\documentstyle[multicol,aps,prl]{revtex}
\begin{document}
\draft
\title{Stretching of polymers in a random three-dimensional flow}
\author{Alexander Groisman and Victor Steinberg}
\address{
Department of Physics of Complex Systems\\
The Weizmann Institute of Science, 76100 Rehovot, Israel}
\date{\today}
\maketitle
\begin{abstract}
Behavior of a dilute polymer solution in a random three-dimensional flow with an average shear 
is studied experimentally. Polymer contribution to the shear stress is found to
be more than two orders of magnitude higher than in a laminar shear flow. The
results indicate that the polymer molecules get strongly stretched by the random 
motion of the fluid.
\end{abstract}
\begin{multicols}{2}
\narrowtext

Behavior of flexible polymer molecules in a solution in various flows is one of the basic
problems of polymer physics \cite{Bird}. Flows that have been studied most extensively 
are simple
shear and unilateral extension. Those flows are rather easy to create in the laboratory
and the experimental data are most straightforward to interpret. As a result of many 
mechanical \cite{Bird,Srid}, optical \cite{Bird,Leal,Muller}, theoretical \cite{Liu,Gennes}
and especially the most recent single molecule investigations \cite{Chu1,Chu2},
the major issues of polymer dynamics in these two types of flow seem to be resolved now.
In contrast to it, there are virtually no experimental data on dynamics of polymer 
molecules in a generic case of a complex flow with a random velocity field. 
In this letter we present results of our measurements of mechanical properties of a dilute
polymer solution in a random three-dimensional (3D) flow. These results directly indicate 
that polymer molecules can become strongly stretched in such a flow. \\
Stretching of polymer molecules manifests macroscopically as growth of optical anisotropy 
of the polymer solution and growth of the stress, $\tau^p$, which is due to polymers 
\cite{Bird}. By definition, $\tau^p=\tau-\tau^s$, where $\tau$ is the total stress in the 
solution, and $\tau^s$ is the stress in the same flow of a pure solvent.\\
Three-dimensional flows with a random velocity field is quite a general class of flows, which
also includes turbulence. Such flows usually occur at high Reynolds number, $Re$, that 
implies large fluid velocity and big size of the tank. The high $Re$ flows are usually
non-homogeneous in space even in statistical sense. Therefore, bulk optical measurements
are difficult to interpret, and so are measurements of mechanical stresses, $\sigma$, at a wall 
of the flow system. In a random, high $Re$ flow the momentum is transferred across the 
fluid not only by the molecular forces contributing to the stress, $\tau$, but also by the 
macroscopic motion of the fluid. The latter produces the Reynolds stress $\rho<v_i v_j>$, 
where $\rho$ is the fluid density, and $\vec{v}$ is fluctuating part of the velocity
$\vec{V}$. Thus, we have a momentum balance equation for the average values of the stresses
$$\sigma=\tau^s+\tau^p+\rho<v_i v_j>, \eqno(1)$$ 
where the right hand side is taken in the bulk of the fluid. That is, in order to calculate 
the average of $\tau^p$, one needs to evaluate the average of $\rho<v_i v_j>$ over 
the whole volume of the flow, which is rather difficult to carry out.\\
A way to surmount the problem of the high Reynolds stress is to create a random flow with a 
low Reynolds number. Such a situation is realized in the case of the elastic turbulence, which 
we have reported recently \cite{Ours}. It is an apparently turbulent flow that can arise in 
polymer solutions as a result of non-linear elastic effects at arbitrarily low $Re$.\\
We carried out our experiments in a swirling flow between two parallel plates. 
Polymer solution was held in a stationary cylindrical cup with a flat bottom 
(lower plate). A coaxial rotating upper plate was just touching the surface of 
the fluid. The cup was mounted on top of a commercial rheometer (AR-1000 of 
TA-instruments) with a normal force transducer. The upper plate was attached to
the shaft of the rheometer, which allowed precise control of its rotation velocity, $\Omega$,
and measurements of the torque, $T$, and the average stress, $\sigma$, applied to it. 
The sidewalls of the cup were made transparent which allowed measurements of the
flow velocity in the horizontal plane by a laser Doppler velocimeter, LDV.
In the first experiment, the radii of the
upper plate and of the cup were $R=30$ mm and $R_2=43.6$ mm, respectively, and the 
distance between the plates was $d=30$ mm. Because of the
big gap and short distance to the side wall, the shear rate profile was rather
non-homogeneous even in a laminar flow. So, we defined the average applied shear rate as 
the ratio between $\sigma$ in a laminar flow, and the fluid viscosity. 
It was $\dot{\gamma}_{av}=2.78\Omega R/d$.
The whole set-up was put into a transparent box, and temperature of fluid was stabilized 
to better than 0.05 $^\circ$C by throughflow of air.\\ 
We used 25 ppm solutions of high molecular weight polyacrylamide 
(M$_{w}$=18,000,000, Polysciences) in viscous sugar syrups. The syrups were made out of
sucrose and sorbitol (corn sugar) in a proportion of 1 : 2, and 1\% of NaCl was
added to stabilize the ionic contents. In the first experiment
the total sugar concentration was 76.3\%, and the solvent viscosity at the
temperature of the experiment, 18 $^\circ$C, was $\eta^s=1.36$ Pa$\cdot$s. 
The polymer part of the solution viscosity, $\eta^p\equiv(\tau-\tau^s)/\dot{\gamma}$, 
was changing at $\dot{\gamma}_{av}$, that we applied, in the range from $0.12\eta^s$ to
$0.08\eta^s$, decreasing with the shear rate. At such low values of $\eta^p/\eta^s$
effect of overlapping of separate polymer molecules is supposed to be minor.\\ 
When flexible polymers get stretched in a shear flow, they also get aligned along the flow
direction \cite{Bird}. This leads to a difference in the normal stresses in the streamwise 
and transverse directions, that for a flow between two rotating plates is 
$N_1=\tau_{\phi\phi}-\tau_{zz}$. Here $(r, \phi, z)$ are cylindrical coordinates.
This first normal stress difference results in a normal force, $F_n$, which pushes the plates
apart (the "hoop stress") \cite{Bird}. So, $N_1$ can be evaluated from 
measurements of $F_n$. For flexible polymer molecules, $N_1$ is connected with the polymer 
relaxation time, $\lambda$, by $N_1=2\eta^p \lambda\dot{\gamma}^2$ \cite{Bird}. For our
solution we measured $\lambda=6.3$ s.\\ 
In the first experiment we evaluated dependence of the stress at the upper plate, $\sigma$, and 
the normal force, $F_n$, on the Weissenberg number, 
$Wi=\lambda\dot{\gamma}_{av}$. (The role of the Weissenberg number in the elastic
turbulence is similar to the role of the Reynolds number in the usual turbulence in 
normal fluids \cite{Ours}.) Simultaneously, one velocity component was measured 
in the center of the set-up. Figure 1a,
curve 1, shows $\sigma$ divided by the stress, 
$\sigma_{lam}$, measured in a laminar flow with the same applied shear rate, as
a function of $Wi$. It resembles very much Fig.2 from Ref.10. The flow at
low $Wi$ is laminar. At $Wi$ of about 5
transition from the laminar flow to the elastic turbulence occurs. It manifests
in growth of the flow resistance and in onset of fluctuations of the fluid 
velocity, Fig.1b. The amplitude of the velocity fluctuations increases with $Wi$. The 
ratio $\sigma/\sigma_{lam}$ increases with $Wi$ as well, reaching a value of about 13 at 
$Wi=22$. The Reynolds number, $Re=\Omega R d\rho/\eta_s$, is only 1.3 at the highest 
$Wi$.\\
As we showed elsewhere \cite{Ours}, the flow of a polymer solution at high $Wi$ (above
the transition) bears all features of developed turbulence. The flow velocity
changes randomly in space and in time, and the fluid motion is excited in a broad range
of spatial and temporal scales. The subject of the current study is stretching of the polymer 
molecules in this random 3D flow. The elastic turbulence itself is driven by
the polymer stresses, which are generated by the stretched polymer molecules. 
However, origin of forces driving a flow is not directly relevant to
the problem of polymer stretching. Extension of polymers can only depend on local 
properties of the flow velocity field along the trajectory of the fluid element,
which contains the polymer molecules.\\ 
In spite of turbulent character of the fluid motion at high $Wi$, the Reynolds stresses 
were so small, that the corresponding term in Eq.(1) could 
be totally neglected. So, at $Wi$=22 characteristic amplitude of the velocity fluctuations,
$V_{rms}$, measured in different points was about 0.7 mm/s (see Fig.1b for $V_{rms}$ 
in the center). 
Thus, the Reynolds stress could be estimated as $\rho V_{rms}^2=7\cdot 10^{-4}$ Pa, 
while the stress $\sigma$ was 91 Pa. Eq. (1) is written in a general
form valid for momentum transfer in the direction perpendicular to a wall. In our case we 
have circular symmetry. So, it is torque, $T$, at the upper plate, which is measured, and it 
is angular momentum, which is transferred from the plate through the liquid to the stationary 
cup. The average flux of angular momentum due to the solvent stress, $\tau^s$, is defined by 
$\eta^s r^2 (\partial\bar{\omega}/\partial r)$ and 
$\eta^s r (\partial\bar{\omega}/\partial \phi)$, where $\bar{\omega}\equiv \bar{V}_{\phi}/r$
is local average angular velocity of the fluid. The turbulent flow changes the distribution
of $\bar{\omega}$ compared to the laminar case, leading to larger gradients of $\bar{\omega}$
near the upper plate \cite{Ours2}. Nevertheless, the boundary conditions on $\omega$ at 
the surfaces of the upper plate and of the cup remain the same. 
So, for our estimates we take the volume average of the flux of the angular momentum due to 
solvent to be the same as in the laminar flow. Then the whole increment in $T$ and 
$\sigma$ should be solely due to growth of average $\tau^p$ in the fluid bulk. Taking 
contribution of the solvent stresses to $\sigma$ to be the same as in the laminar flow, we 
get for the contribution of the polymer stresses to $\sigma$ the curve 2 in Fig.1a. This curve
gives the ratio of the average shear stresses due to polymers in the turbulent and the
laminar flows, $\tau^p/\tau^p_{lam}$. This ratio reaches a value of 170, 
which is an evidence of strong stretching of the polymers in the turbulent flow.\\
Dependencies of average $N_1$ on $Wi$ measured in the elastic turbulent flow and in a 
laminar flow are shown in Fig.1c. One can learn that $N_1$ in the turbulent flow becomes 
about an order of magnitude higher than in the laminar flow with the same 
$\dot{\gamma}_{av}$. This is another evidence of stretching of the polymers by the random
3D fluid motion. \\
A specific feature of the elastic stresses is that they do not turn to zero immediately
after the fluid motion stops, but rather decay with their characteristic relaxation time, 
$\lambda$ \cite{Bird}. That is how they can be clearly distinguished from the viscous stresses, 
which decay instantaneously. So, a way to independently examine the origin of stresses applied
to the upper plate is to stop its rotation and to measure decay of the stresses. This
was the objective of our second experiment, Fig.2. It required a higher polymer relaxation time. 
Therefore, the concentration of sugars was increased to 80.3\%, that gave $\eta^s$ 
of about 7.2 Pa$\cdot$s and $\lambda$ of about 30 sec at the temperature of the experiment, 
15 $^\circ$C. The size of the set-up was reduced by a factor of 2, $R=15$ mm, 
$R_2=21.8$ mm, $d=15$ mm. In this small set-up the dependence of $\sigma$ on 
$Wi$ for the polymer solution from the first experiment was the same as 
in Fig.1a (see also Ref.10). The characteristic torques were 8 times smaller, however,
that enabled a sharp stop of the upper plate.\\
Rotation of the upper plate was started
abruptly at angular velocity $\Omega=0.07$ s$^{-1}$ corresponding to $Wi\approx6$ and
$Re\approx 3\cdot10^{-3}$. The 
rotation continued for a while, and then, at a time moment taken as zero, $\Omega$ was 
abruptly brought to zero. In the first run, curve 1 in Fig.2b, the time of the rotation was 
short, about 115 seconds. It was just enough for the polymers to get properly stretched by the 
primary shear and for the polymer stress to reach its laminar flow value, $\tau^p_{lam}$.
So, by the time the rotation was stopped at $t=0$, the major part of torque was due to the
solvent, $T^s\approx 10$ $\mu$N$\cdot$m, and it relaxed almost immediately. 
(Characteristic viscous diffusion time is very small, $d^2\rho/\eta^s=0.04$ s.) This torque 
value, $10$ $\mu$N$\cdot$m, was also measured just after rotation had been started 
and polymers had not had time to get stretched yet. Extrapolation of the curve 1
in Fig.2b to $t=+0$ gives the value of the torque due to polymers in the laminar flow, 
$T^p_{lam}=0.85$ $\mu$N$\cdot$m.\\
In the second run the time of rotation, about 4500 sec, was long enough for the
transition to developed elastic turbulence to complete. The
torque reached its saturated value, $T=82.5$ $\mu$N$\cdot$m, corresponding to applied $Wi$.
From Fig.2a one can see that the fluid velocity was fluctuating
during the time, that the torque increased. As a result of the turbulent 
fluid motion, the slowly relaxing part of 
the torque increased by almost two orders of magnitude. One can learn from the inset that
the relaxing torque can be reasonably extrapolated to $72.5$ $\mu$N$\cdot$m at $t=+0$. Thus, 
the immediately relaxing part of the torque is again due to the solvent shear stresses, 
$T^s \approx 10$ $\mu$N$\cdot$m, and the results of these relaxational measurements are quite 
consistent with the suggestion, that the whole increment in the torque is due to growth of
the polymer contribution. If inertial effects were important, they would obviously lead to 
torques of the opposite sign. 
The polymer torque $T^p=72.5$ $\mu$N$\cdot$m at $t=0$ means average increase of $\tau^p$
by 85 times compared with the laminar flow.\\
Our measurements show that as a result of a secondary random 3D flow, superimposed on the 
primary applied shear, the polymer contribution to shear stress, $\tau^p_{z\phi}$, can become 
as much as 170 times larger. If linear elasticity of the polymer molecules is assumed,
the stress is related to the polymer extension via
$\tau^p_{z\phi} \sim <R_z R_{\phi}>$, where $\vec{R}$ is the vector connecting edges of 
a polymer chain \cite{Bird}. Then the factor of additional extension of the polymer molecules 
due to the random 3D flow is about 13. In fact, measurements of the stress in the 
turbulent regime can be viewed as a mechanical test, where shear is applied to a polymer 
solution, which flows turbulently with zero average shear rate. Then, one can say that
polymer contribution to viscosity, defined as $\eta^p=\tau^p_{z\phi}/\dot{\gamma}_{av}$,
increases by more that two orders of magnitude because of the turbulent flow. In theories
of polymer dynamics $\eta^p$ is always connected with an effective volume occupied
by the polymer chains \cite{Bird}. Thus, our experimental results suggest that the 
effective volume can significantly increase, and polymer molecules can get strongly stretched 
in a random 3D flow with zero averages.\\
Dynamics of a polymer molecule in a random 3D flow were first considered by Lumley \cite{Lumley},
and have been revised recently \cite{Lebed,Cher}. It is suggested that the flow is always 
homogeneous on the scale of a polymer molecule, so that velocity field in some vicinity 
of a molecule with the center at 
$\vec{r}_0$ is given by $\vec{V}=\vec{V}(\vec{r}_0)+\kappa\cdot(\vec{r}-\vec{r}_0)$. 
Possible complex structure properties of the flow at larger scales are not important
for the issue of the polymer stretching. It only depends on statistics of the tensor 
of rate of deformation, $\kappa$, which varies randomly in time and space.
If the flow is truly 3D, $\kappa$ always has an eigenvalue with a positive real part, 
so that there is a direction along which pure extension occurs \cite{Leal,Lumley}.
The direction and the rate of extension change randomly as a fluid element is rotating 
in the flow and moving along its trajectory. Nevertheless, if $\kappa$ remains correlated 
within finite time intervals, the overall statistical result of such random motion will 
be exponential divergence of two closely spaced material points. In a turbulent flow an 
estimate for the correlation time of $\kappa$ is given by the inverse of the velocity gradients
themselves. Then the average Lyapunov exponent, $a$, for divergence of two material points is 
given by the rms of longitudinal velocity gradients
with a prefactor of order unity. Thus, in the statistical sense, a random 3D flow acts 
as an extensional flow with $\partial V_x/\partial x=a$, where the direction, $x$,
of the maximal extension of a fluid element is changing randomly in time and space 
\cite{Lumley,Lebed}. 
In such a flow polymer molecules should become vastly stretched, if the condition 
$a\lambda>1/2$ is fulfilled \cite{Gennes,Chu1,Lumley,Lebed}. In fact, strong polymer stretching 
in a turbulent flow is expected to occur even at smaller average $a$ \cite{Lebed}.\\
A probability distribution function (PDF) of a longitudinal velocity
gradient, $\partial V_{\phi}/(r\partial\phi)$,
for $Wi=12$ in the big set-up (first experiment) is shown in Fig.3. 
The rms of the distribution is 0.043 
s$^{-1}$, which gives a reasonable value of 0.25, when multiplied by $\lambda=6.3$ s. In
fact, relaxation of polymer molecules is a complex process, which involves a whole range of 
times. We measured relaxation of $\tau^p$ in our polymer solution after a 
sudden stop of a stationary shear flow with $Wi=6$. The apparent relaxation 
time, defined as $(\partial ln(\tau^p)/\partial t)^{-1}$, increased from 3 s to 40 s as 
the stress was decaying (see also Fig.2b). Further, the PDF in Fig.3 has pronounced exponential 
tails, so that flow events with large rate of extension occur rather often. Therefore,
we can conclude that the observed significant stretching of the polymers
is quite consistent with the theoretical predictions. We believe that a similar kind
of polymer stretching should occur in the usual, high $Re$, turbulent flows, when
the velocity gradients are sufficiently large compared to $\lambda$. 

We gratefully acknowledge E. Balkovsky, G. Falkovich, A. Fouxon and V. Lebedev for numerous
helpful discussions. The work was partially supported by the Minerva Center for Nonlinear 
Physics of Complex Systems and by a Research Grant from the Henri Gutwirth Fund.

\begin{figure}
\caption{Dependence of different parameters of flow on $Wi$. $Wi$
was raised by 9\%/min, 2.4\%/min and 7\%/min in the ranges [0.35, 2.1],
[2.1, 8.7] and [8.7, 25], respectively. {\bf a)} The ratio of the total stresses, 
$\sigma/\sigma_{lam}$, curve 1. The
ratio of the polymer contribution to the stress, 
$\tau^p/\tau^p_{lam}$, curve 2. {\bf b)} $V_{rms}$ in the center of the set-up, curve 1. 
Instrumental noise level in a laminar flow is shown
for comparison, curve 2. {\bf c)} Average first normal stress difference, $N_1$, in the
turbulent flow, curve 1, and in a laminar flow between two plates, curve 2.}

\label{figa}
\end{figure}

\begin{figure}
\caption{Dependence of different parameters of flow on time, $t$. The time
axis is compressed by a factor of 100 for $t<0$. {\bf a)} Flow velocity in the center
corresponding to the curve 2 below. {\bf b)} The torque applied to the upper plate.
{\it Inset:} curve 2 in a vicinity of $t=0$.}

\label{figb}
\end{figure}

\begin{figure}
\caption{PDF for the longitudinal velocity gradients, 
$\partial V_{\phi}/(r\partial\phi)$, estimated according to the Taylor 
hypothesis, with smoothing over about 1.3 mm. $V_{\phi}$ was 
measured at $z=3.75$ mm from the upper 
plate, at $r=15$ mm, with time intervals $\Delta t = 0.44$ sec. The differences between 
the consecutive velocity readings were divided by $\Delta t$ and by 
$\bar{V}_{\phi}=2.93$ mm/s. The thin line is a Gaussian fit.}
\label{figc}
\end{figure}

\end{multicols}

\end{document}